\documentclass[%
 reprint,
superscriptaddress,
 amsmath,amssymb,
 aps,
 prd,
floatfix,
longbibliography,
eprint,
]{revtex4-1}

\usepackage[utf8]{inputenc}
\usepackage{commath}
\usepackage{esdiff} 
\usepackage{nicefrac}
\usepackage{tabularx}
\usepackage{graphicx}
\usepackage[caption=false]{subfig}
\usepackage{dcolumn}
\usepackage{bm}
\usepackage[colorlinks]{hyperref}
\usepackage{siunitx} 
\usepackage{xcolor}
\usepackage[utf8]{inputenc}
\usepackage{etoolbox} 
\usepackage[capitalise,nameinlink]{cleveref} 
\Crefname{equation}{Eq.}{Eqs.}
\Crefname{figure}{Fig.}{Figs.}
\Crefname{tabular}{Tab.}{Tabs.}
\makeatletter
\appto{\appendix}{%
  \@ifstar{\def\theequation@prefix{A.}}%
          {}%
}
\makeatother

\DeclareSIUnit\parsec{pc}
\DeclareSIUnit\efolds{e\text{-}folds}
\DeclareSIUnit\planckmass{m_\mathrm{p}}

\begin{document}

\preprint{APS/123-QED}

\title{Constraining the kinetically dominated Universe}

\author{L.~T.~Hergt}
 \email{lh561@mrao.cam.ac.uk}
\author{W.~J.~Handley}%
 \email{wh260@mrao.cam.ac.uk}
\affiliation{%
 Astrophysics Group, Cavendish Laboratory, J.~J.~Thomson Avenue, Cambridge, CB3~0HE, UK
}%
\affiliation{%
 Kavli Institute for Cosmology, Madingley Road, Cambridge, CB3~0HA, UK
}%
\author{M.~P.~Hobson}%
 \email{mph@mrao.cam.ac.uk}
\affiliation{%
 Astrophysics Group, Cavendish Laboratory, J.~J.~Thomson Avenue, Cambridge, CB3~0HE, UK
}%
\author{A.~N.~Lasenby}%
 \email{a.n.lasenby@mrao.cam.ac.uk}
\affiliation{%
 Astrophysics Group, Cavendish Laboratory, J.~J.~Thomson Avenue, Cambridge, CB3~0HE, UK
}%
\affiliation{%
 Kavli Institute for Cosmology, Madingley Road, Cambridge, CB3~0HA, UK
}%

\date{\today}

\begin{abstract}
We present cosmological constraints from Planck 2015 data for a universe that is kinetically dominated at very early times. We perform a Markov chain Monte Carlo analysis to estimate parameters and use nested sampling to determine the evidence for a model comparison of the single-field quadratic and Starobinsky inflationary models with the standard $\Lambda$CDM cosmology. In particular we investigate how different amounts of inflation before and after horizon exit affect the primordial power spectrum and subsequently the power spectrum of the cosmic microwave background. We find that the model using kinetically dominated initial conditions for inflation performs similarly well in terms of Bayesian evidence as a model directly starting out in the slow-roll phase, despite having an additional parameter. The data show a slight preference for a cutoff at large scales in the primordial and temperature power spectra.

\end{abstract}

\maketitle

\section{\label{sec:intro}Introduction}
Inflation was first introduced in the 70s and 80s (see~\cite{Starobinsky1979,Guth1981,Linde1982} for some of the original papers and section~2 of~\cite{Planck2013inflation} for a more extensive introduction) and plays an important role in today's standard model of cosmology ($\Lambda$CDM). Besides solving issues such as the horizon and flatness problems, it provides a mechanism for generating primordial perturbations that can serve as seeds for the formation of cosmic structure, which in turn generate the observed temperature anisotropies in the cosmic microwave background (CMB)~\cite{Mukhanov1992}. 

Typically, a slow-roll~(SR) inflation model is assumed, whereby the kinetic energy of a single scalar field~$\phi$ is dominated by its potential~$V(\phi)$ and hence the inflaton ``slowly rolls down'' the potential. Generically, the slow-roll condition is an attractor solution so even from a position in phase space where slow-roll is not satisfied, the inflaton will rapidly lose speed and approach a slow-roll regime~\cite{Belinsky1985,Linde1985,Boyanovsky2006,Boyanovsky2006a,Destri2008,Destri2010,Handley2014,KineticNote,Hergt2018}.

High-precision measurements of the CMB, first through WMAP\cite{WMAP0} then through Planck~\cite{Planck2013overview,Planck2015overview}, have significantly contributed to the success of the standard $\Lambda$CDM model of cosmology. Nonetheless, the data also revealed features in the CMB angular power spectrum hinting at potential additional physics~\cite{WMAP1,WMAP2,WMAP3,Mortonson2009}. These features include the low-multipole lack of power and a small dip at multipoles of approximately 20--25. These features may be caused by corresponding features in the primordial power spectrum (PPS), which recently has led to many investigations of PPS with a cutoff~\cite{Contaldi2003,Destri2008,Destri2010,Ramirez2012,Ramirez2012a,Scacco2015,Planck2015inflation,DaCosta2017}. 

In this paper, we look in more detail into the effects of a kinetically dominated~(KD) early universe which is shown to emerge generically from an initial singularity under rather broad assumptions in~\cite{Handley2014,KineticNote}. This is particularly relevant for inflationary potentials that have an upper limit in the inflaton range of interest, such as plateau or hilltop potentials~\cite{Hergt2018}. Another way of motivating KD is through the ``just enough inflation'' scenario~\cite{Ramirez2012,Ramirez2012a}. We show how KD initial conditions result in oscillations and a cutoff towards large scales in the PPS and consequently also in the CMB angular power spectrum. We show how these features depend mainly on the amount of inflation happening before or after horizon exit of a given mode~$k$ and perform a Markov chain Monte Carlo (MCMC) analysis to estimate cosmological parameters given KD initial conditions and compare the evidences for the different models. 

We start out by summarising the inflationary background evolution in \cref{sec:background}, and by introducing two inflationary potentials, the quadratic and the Starobinsky potential, which we will use throughout this paper. In \cref{sec:kd} we review the kinetic dominance regime that provides us with the initial conditions for the numerical integration of the inflaton equations of motion and the mode equations for the primordial perturbations, which lead us to the analyses of the PPS in \cref{sec:pps} and the CMB angular power spectrum in \cref{sec:cmb}. Finally, in \cref{sec:mcmc} we present the results from our MCMC analysis and conclude in \cref{sec:conclusions}.

\section{\label{sec:background}Background Evolution during Kinetic Dominance}

\begin{figure*}[tb]
		\centering
		\includegraphics[]{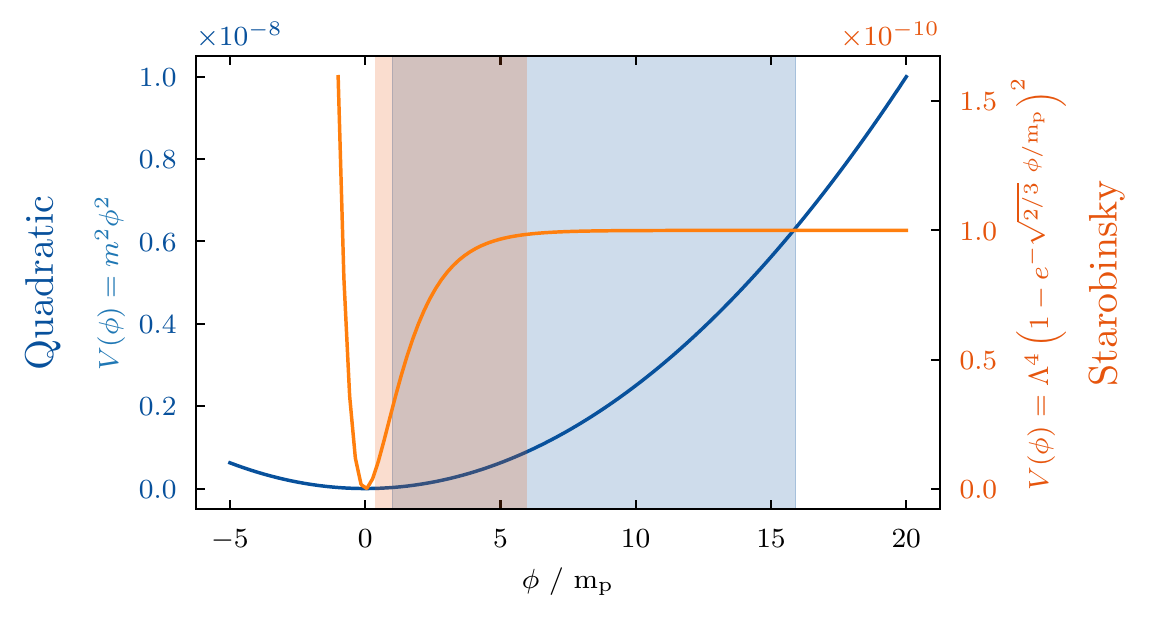}
	    \caption{Chaotic (blue) and Starobinsky (orange) potentials as functions of the inflaton field~$\phi$, where $m=\SI{5e-6}{\planckmass}$ for the Quadratic potential and $\Lambda^2=\SI{e-5}{\planckmass\squared}$ for the Starobinsky potential. The shaded regions mark the start and end of inflation in the case of kinetic dominance initial conditions.}
    \label{fig:potentials}
\end{figure*}

We focus on single-field inflationary models as determined by an inflaton field~$\phi(t)$ in a spatially flat universe. Assuming the inflaton dominates all other species early in the history of the Universe, the background dynamics are governed by the Friedmann and continuity equations for the inflaton 
\begin{subequations}
\begin{align}
	\label{eq:background1} 
	H^2 &= \frac{1}{\SI{3}{\planckmass\squared}} \left(\frac{1}{2} \dot\phi^2 + V(\phi)\right), \\
	\label{eq:background2}
    \dot H &= - \frac{1}{\SI{2}{\planckmass\squared}} \dot\phi^2,  \\
    \label{eq:eom}
    \ddot\phi &+ 3 H \dot\phi + V'(\phi) = 0 , 
\end{align}
\end{subequations}
where a dot denotes differentiation with respect to cosmic time, $\dot f\equiv \tod{f}{t}$. For convenience we set $c=\hbar=1$ and use the reduced Planck mass $\smash{\si{\planckmass}=\sqrt{\frac{\hbar c}{8\pi G}}}$. 

Inflation is defined as a positive acceleration of the scale factor $\ddot a > 0$, or equivalently as a shrinking comoving Hubble horizon $\frac{\dif }{\dif t} \left(\frac{1}{aH}\right) < 0$. Using \cref{eq:background1,eq:background2,eq:eom} we can recast this condition for inflation in terms of the inflaton field~$\phi$
\begin{equation}
	\dot\phi^2 < V(\phi) ,
\end{equation}
or in terms of the equation-of-state parameter~$w\equiv\frac{p}{\rho}$ relating pressure~$p$ and energy density~$\rho$ of the inflaton
\begin{equation}
	\label{eq:eos}
	w_\phi = \frac{p_\phi}{\rho_\phi} = \frac{\frac{1}{2} \dot\phi^2 - V(\phi)}{\frac{1}{2} \dot\phi^2 + V(\phi)} < - \frac{1}{3} .
\end{equation}
The amount of inflation from some time~$t$ to the end of inflation~$t_\mathrm{end}$ can be measured in terms of the number of \si{\efolds} of the scale factor~$a(t)$
\begin{equation}
	N(a) \equiv \ln\Big(\frac{a_\mathrm{end}}{a}\Big) ,
\end{equation}
where $a_\mathrm{end}=a(t_\mathrm{end})$.

\subsection{Potentials}

To perform numerical integrations of the background dynamics in \cref{eq:background1,eq:background2,eq:eom} we have focused on two specific potentials in particular: the quadratic potential and the Starobinsky potential shown in \cref{fig:potentials}.

\subsubsection{Quadratic potential}

The quadratic potential is defined by
\begin{equation}
	\label{eq:m2phi2}
	V(\phi) = m^2 \phi^2 ,
\end{equation}
where $m$ is the mass of the inflaton field. This quadratic potential is often defined with a multiplicative factor~$\frac{1}{2}$, omitted here for reasons of compatibility with other power law potentials.  Though disfavoured by the Planck data, we are considering the quadratic potential here as the conceptually simplest implementation of a single scalar inflaton field. 
Using the slow-roll~(SR) approximation $\dot\phi^2 \ll V(\phi)$ we can predict the spectral index and the tensor to scalar ratio to be
\begin{equation}
	\label{eq:m2phi2_ns_r}
	n_\mathrm{s} \approx 1 - \frac{2}{N_\ast} , \qquad
    r \approx \frac{8}{N_\ast} , 
\end{equation}
where~$N_\ast$ is the observable amount of inflation from horizon exit of a given pivot scale~$k_\ast$ to the end of inflation.
Thus for $N_\ast = \SI{55}{\efolds}$ we expect $n_\mathrm{s} \approx 0.964$ and $r \approx 0.145$. 

\subsubsection{Starobinsky potential}

The Starobinsky potential is the potential representation in the Einstein frame of an $(R+R^2)$ modified theory of gravity first proposed by~\cite{Starobinsky1980} and is given by
\begin{equation}
	\label{eq:starobinsky}
	V(\phi) = \Lambda^4 \left[ 1 - \exp\left(-\sqrt{\frac{2}{3}} \frac{\phi}{\si{\planckmass}}\right) \right]^2 . 
\end{equation} 
Unlike quadratic inflation, the Starobinsky model gives rise to a low tensor-to-scalar ratio~$r$, as is preferred by current data~\cite{Planck2015inflation}. In the same manner as quadratic inflation, we can determine the spectral index and the tensor to scalar ratio using the slow-roll approximation
\begin{equation}
	\label{eq:starobinsky_ns_r}
	n_\mathrm{s} \approx 1 - \frac{2}{N_\ast} , \qquad 
    r \approx \frac{12}{N_\ast^2} .
\end{equation}
Thus for $N_\ast = \SI{55}{\efolds}$ we expect $n_\mathrm{s} \approx 0.964$ and $r \approx 0.004$.

\subsection{\label{sec:kd}Kinetic Dominance initial conditions}

\begin{figure}[!tbp]
	\centering
	\includegraphics[width=\columnwidth]{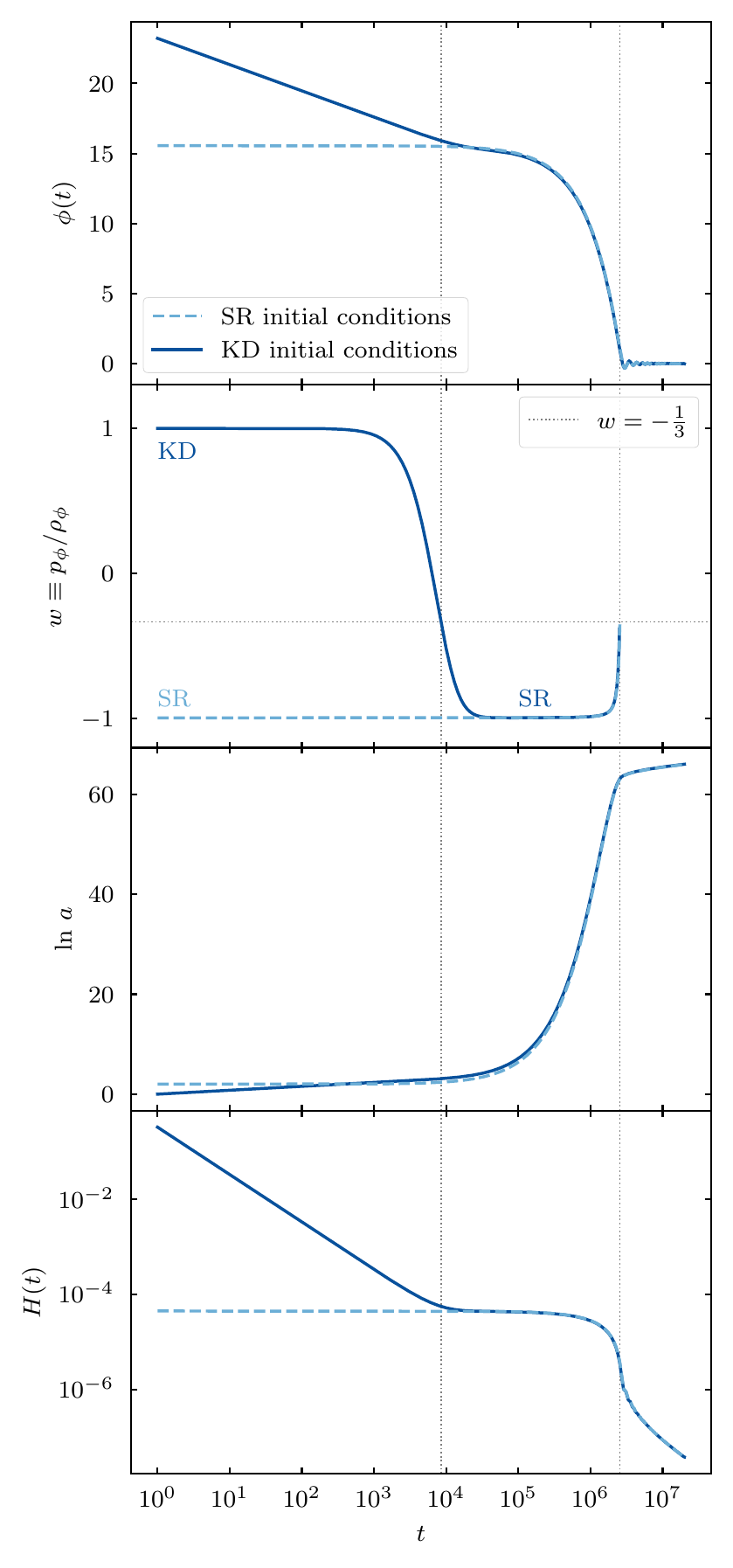}
	\caption{\label{fig:background}
Evolution of the inflaton field~$\phi(t)$, equation-of-state parameter~$w_\phi(t)$, scale factor~$a(t)$, and Hubble parameter~$H(t)$ respectively for the quadratic potential from \cref{eq:m2phi2}. The inflaton mass was taken to be $m=\SI{5e-6}{\planckmass}$. The light dashed line starts out directly in the slow-roll~(SR) regime, whereas the dark solid line starts out during kinetic dominance~(KD) and then later joins the SR attractor. The initial conditions were set such that $N_\mathrm{tot}=\SI{60}{\efolds}$ of inflation are produced. The equation-of-state parameter~$w_\phi$ is useful in determining the start and end of inflation in the KD case (dotted lines).}
\end{figure}

\begin{figure*}[tb]
	\centering
	\includegraphics[]{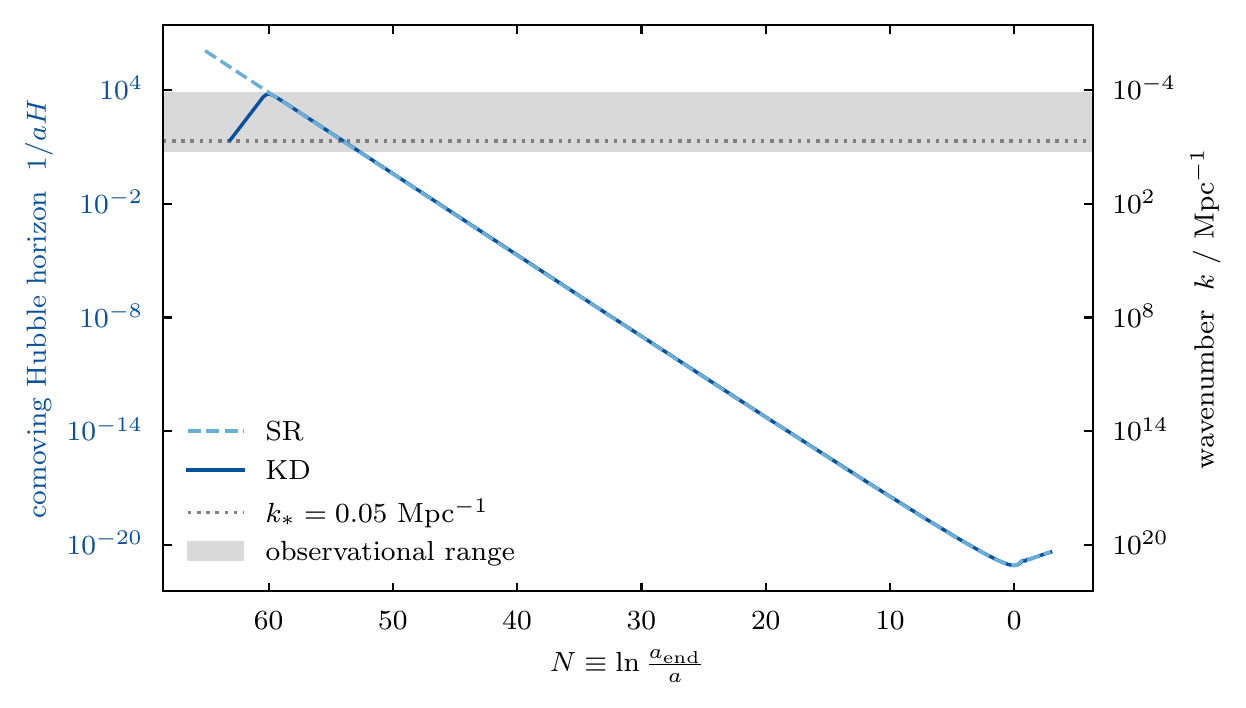}
	\caption{\label{fig:hubble} 
The comoving Hubble horizon is plotted here as a function of~$N$. For the dashed line slow-roll~(SR) initial conditions were used and for the solid line kinetic dominance~(KD) initial conditions. Note how the comoving Hubble horizon falls monotonically for the SR solution, whereas it has a local maximum at the start of inflation for the KD solution, which is not evident in \cref{fig:background}. Thus, in the KD case, there are scales $k^{-1} > (aH)^{-1}$ that were never within the comoving Hubble horizon.}
\end{figure*}

The initial conditions for the integration of the background \cref{eq:background1,eq:background2,eq:eom} are usually chosen according to the slow-roll~(SR) regime, satisfying
\begin{equation}
	\dot\phi^2 \ll V(\phi) .
\end{equation}
However, we do not need to place ourselves (somewhat artificially) directly into the period of SR inflation. As observed previously~\cite{Linde1985,Belinsky1985}, the expansion of the Universe acts as a damping term in the equation of motion~(\ref{eq:eom}). This means the SR condition is an attractor solution, such that no matter where we start out in the $(\phi, \dot\phi)$~phase-space we will end up on the SR attractor (provided one assumes an appropriate inflationary potential). 
Indeed, {\NoHyper\citeauthor{Handley2014}\endNoHyper}~\citep{Handley2014,KineticNote} show under broad assumptions that classical inflationary universes generically emerge from an initial singularity ($a \rightarrow 0$) with the kinetic energy of the inflaton dominating its potential energy~\cite{Handley2014,KineticNote}, which we will refer to as kinetic dominance~(KD)
\begin{equation}
    \label{eq:KD}
	\dot\phi^2 \gg V(\phi) .
\end{equation}
In a recently submitted paper~\cite{Hergt2018}, we make a case for kinetically dominated initial conditions for inflation through a $(\phi, \dot\phi)$~phase-space exploration. This is particularly relevant in cases where the potential is bounded from above, e.g.\ hilltop or plateau potentials.

In the KD limit we can use the first terms of a series expansion of the background variables to generate a set of initial conditions for a sufficiently early starting time~$t_0$ of the numerical integration
\begin{subequations}
\begin{align}
	\phi_0 &\equiv \phi(t_0) = \phi_\mathrm{p} - \SI[parse-numbers=false]{\sqrt{\frac{2}{3}}}{\planckmass} \ln t_0 , 
    \label{eq:phi0} \\
    \dot\phi_0 &\equiv \dot\phi(t_0) = - \sqrt{\frac{2}{3}} \frac{\si{\planckmass}}{t_0} , 
    \label{eq:dphi0} \\
    H_0 &\equiv H(t_0) = \frac{1}{3 t_0} , 
    \label{eq:h0} \\
    a_0 &\equiv a(t_0) = a_\mathrm{p} \left( \frac{t_0}{t_\mathrm{p}} \right)^{1/3} , 
    \label{eq:a0}
\end{align}
\end{subequations}
where~$a_\mathrm{p}$ and~$t_\mathrm{p}$ can be set to unity as the exact value does not matter here due to rescaling symmetries~\cite{Handley2014,KineticNote}. $\phi_\mathrm{p}$ controls the total number of \si{\efolds} of inflation $N_\mathrm{tot}=N(a_\mathrm{start})$, i.e.\ from the start of inflation $a_\mathrm{start}=a(t_\mathrm{start})$ to its end.

\Cref{fig:background} shows the evolution of the background variables~$\phi$, $\ln a$, $w_\phi$, and~$H$ respectively, integrated using both SR and KD initial conditions and using the chaotic potential from \cref{eq:m2phi2}. For this figure, the initial conditions were set at the cosmic time $t_0=1$ and chosen such that $N_\mathrm{tot}=\SI{60}{\efolds}$ are produced during inflation. 
For comparison, the end of inflation in the SR case was shifted such that it matches the KD case. The inflaton mass of $m=\SI{5e-6}{\planckmass}$ was chosen to produce an amplitude~$A_\mathrm{s}$ of the primordial power spectrum close to the observational value. In all cases we see how the evolution begins differently depending on whether SR or KD initial conditions were chosen, but eventually the KD solution converges towards the SR solution. 

To distinguish between the different regimes it is useful to look at the equation-of-state parameter~$w_\phi$ for the inflaton field and comparing with \cref{eq:eos}
\begin{align}
	w_\phi 
    \left\{
    \begin{aligned}
    	&\approx \hphantom{-}1 && \text{kinetic dominance}, &  \dot\phi^2 \gg V(\phi) ,\\
        &> -\tfrac{1}{3}\vphantom{\dot\phi^2} && \text{no inflation} , & \\
        &< -\tfrac{1}{3}\vphantom{\dot\phi^2} && \text{(fast-roll) inflation} , & \\
        &\approx -1 && \text{slow-roll inflation}, &  \dot\phi^2 \ll V(\phi) .
    \end{aligned}
    \right .
\end{align}
The equation-of-state parameter~$w_\phi$ illustrates how in the SR case we directly start out in the inflationary epoch, whereas for the KD case we can specify a start and end point of inflation where~$w_\phi$ crosses the $\nicefrac{-1}{3}$~mark. For reasons of clarity, the evolution of~$w_\phi$ was cut off at the end of inflation, after which it starts oscillating rapidly.

\Cref{fig:hubble} shows the evolution of the comoving Hubble horizon as a function of the logarithm of the scale factor.  As expected it shrinks during inflation. However, during KD the comoving Hubble horizon initially grows until the onset of inflation where it meets the SR solution and starts decreasing. Thus, in a universe initially going through a phase of KD there exists a maximum to the comoving Hubble horizon and consequently there are very large scales that have never been within the horizon before the start of inflation. 

\begin{figure*}[tb]
	\centering
	\subfloat{\includegraphics[width=\columnwidth]{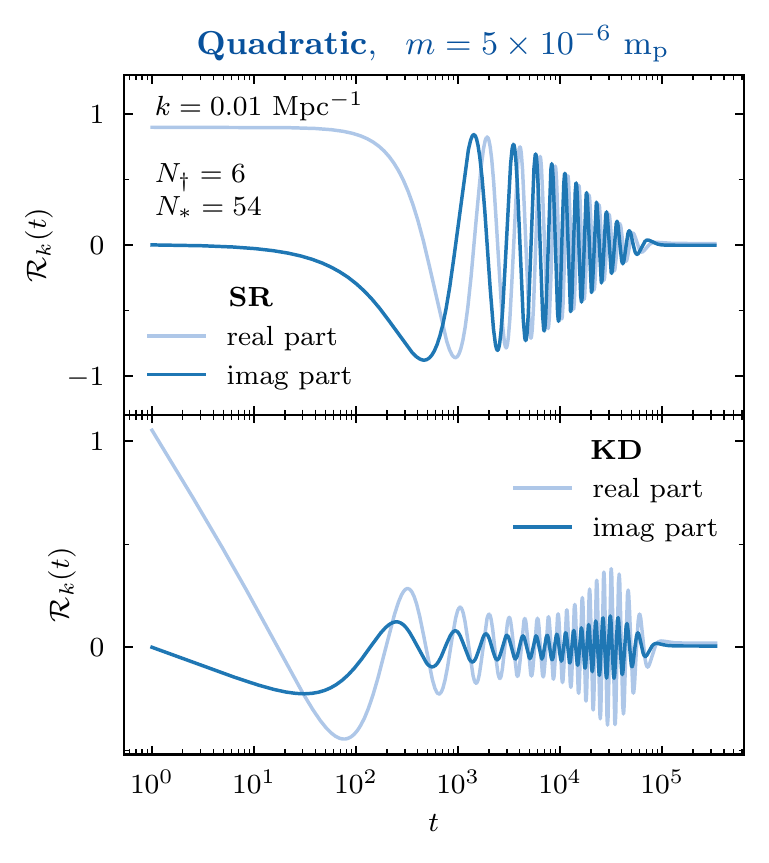}}
    \hfill
    \subfloat{\includegraphics[width=\columnwidth]{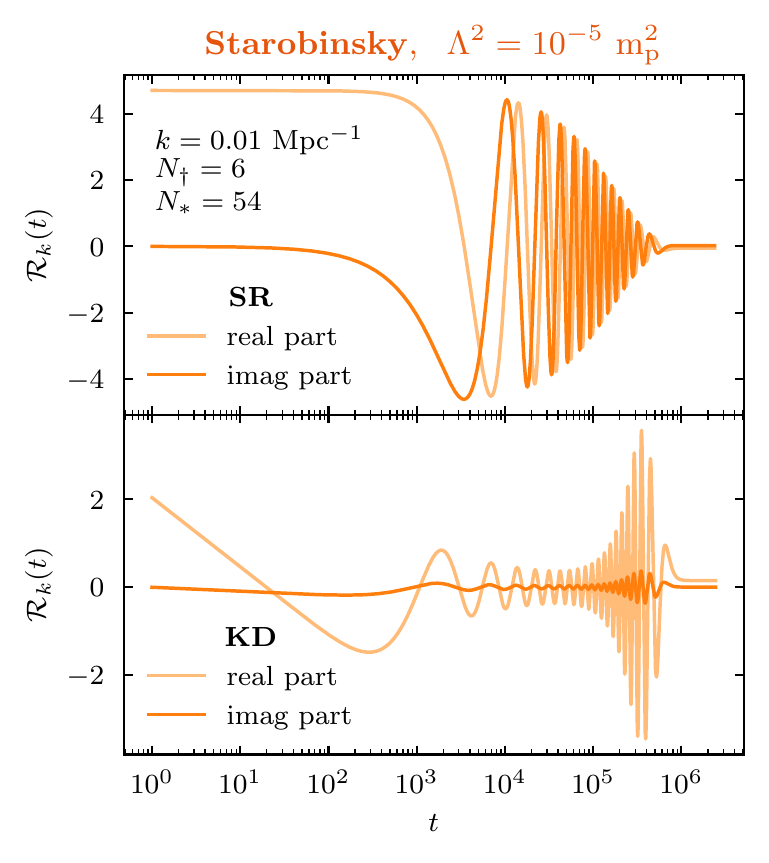}}
    \caption{\label{fig:rk} Evolution of the primordial curvature perturbations~$\mathcal{R}_k(t)$ from \cref{eq:scalarmodes} for the given mode $k=\SI{0.01}{\per\mega\parsec}$ for quadratic inflation on the left with an inflaton mass of $m=\SI{5e-6}{\planckmass}$ and for Starobinsky inflation with an amplitude of $\Lambda^2=\SI{e-5}{\planckmass}$. The background variables were set up using slow-roll~(SR) initial conditions in the top plots and using kinetic dominance~(KD) initial conditions in the bottom plots such that a total number of $N_\mathrm{tot}=\SI{60}{\efolds}$ were produced. In terms of the quantities defined in \cref{sec:efolds}, they are split into $N_\dagger=\SI{6}{\efolds}$ \emph{before} and $N_\ast=\SI{54}{\efolds}$ \emph{after} horizon exit of the pivot scale $k_\ast=\SI{0.05}{\per\mega\parsec}$.}
\end{figure*}

\section{\label{sec:pps}Primordial Power Spectrum}

\begin{figure*}[tb]
  \centering
  \includegraphics[]{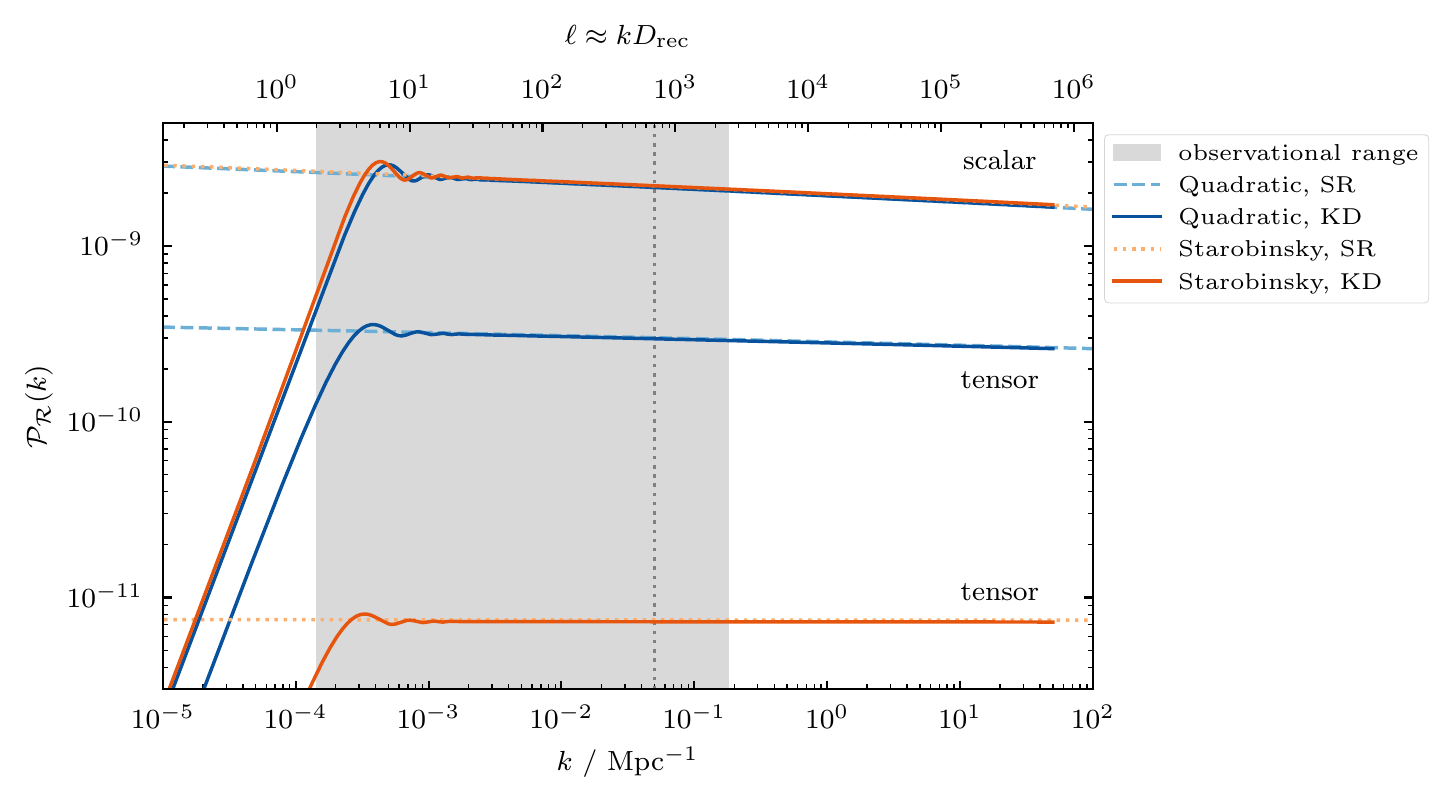}
  \caption{\label{fig:PPSnumeric} Primordial Power Spectra: The upper four lines correspond to the scalar power spectra, i.e.\ from primordial curvature perturbations. The lower four lines correspond to the tensor power spectra, i.e.\ from gravitational waves. On small scales the power spectra from slow-roll~(SR) and from kinetic dominance~(KD) initial conditions agree well with one another and show the characteristic power-law behaviour. Towards larger scales the SR power spectra continue along the power-law slope whereas the KD power spectra start oscillating and eventually show a cutoff. These very large scales are the ones that were never within the comoving Hubble horizon in an initially kinetically dominated universe (cf.\ \cref{fig:hubble}).  The dotted vertical line marks the pivot scale $k_\ast=\SI{0.05}{\per\mega\parsec}$, used for the calculation of spectral index~$n_\mathrm{s}$ and tensor-to-scalar ratio~$r$ as well as for the calibration of the $k$-axis. The parameters governing~$n_\mathrm{s}$, $r$ and the cutoff position were set to the best-fit values from the MCMC analysis in \cref{sec:cmb}. The shaded region gives a rough estimate of the observational window in the CMB power spectrum.}
\end{figure*}

\begin{figure*}[tb]
    \subfloat[PPS for varying~$N_\ast$ at a fixed $N_\dagger=\SI{6}{\efolds}$.]{\includegraphics[width=\columnwidth]{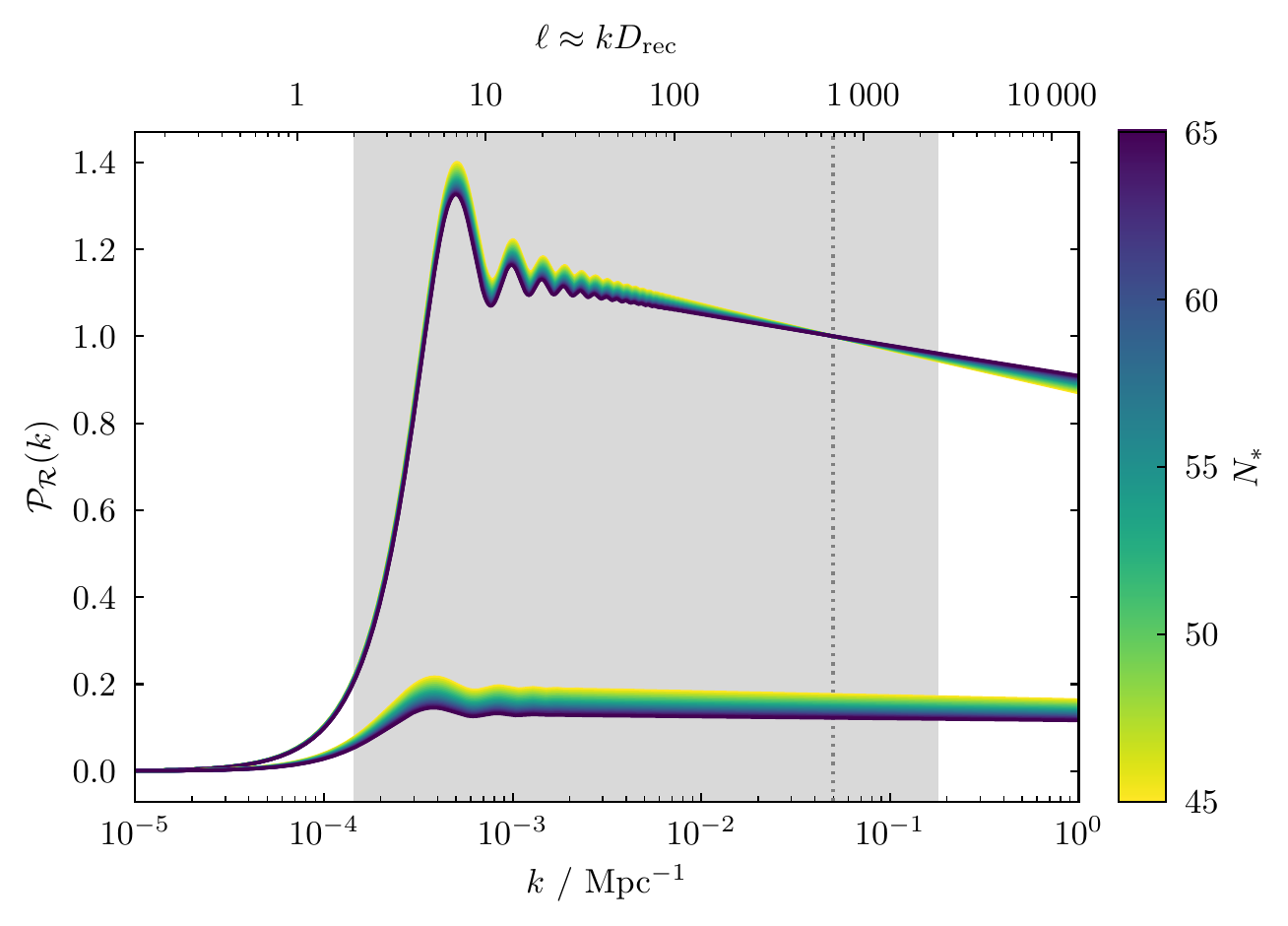}}
    \hfill
    \subfloat[PPS for varying~$N_\dagger$ at a fixed $N_\ast=\SI{55}{\efolds}$.]{\includegraphics[width=\columnwidth]{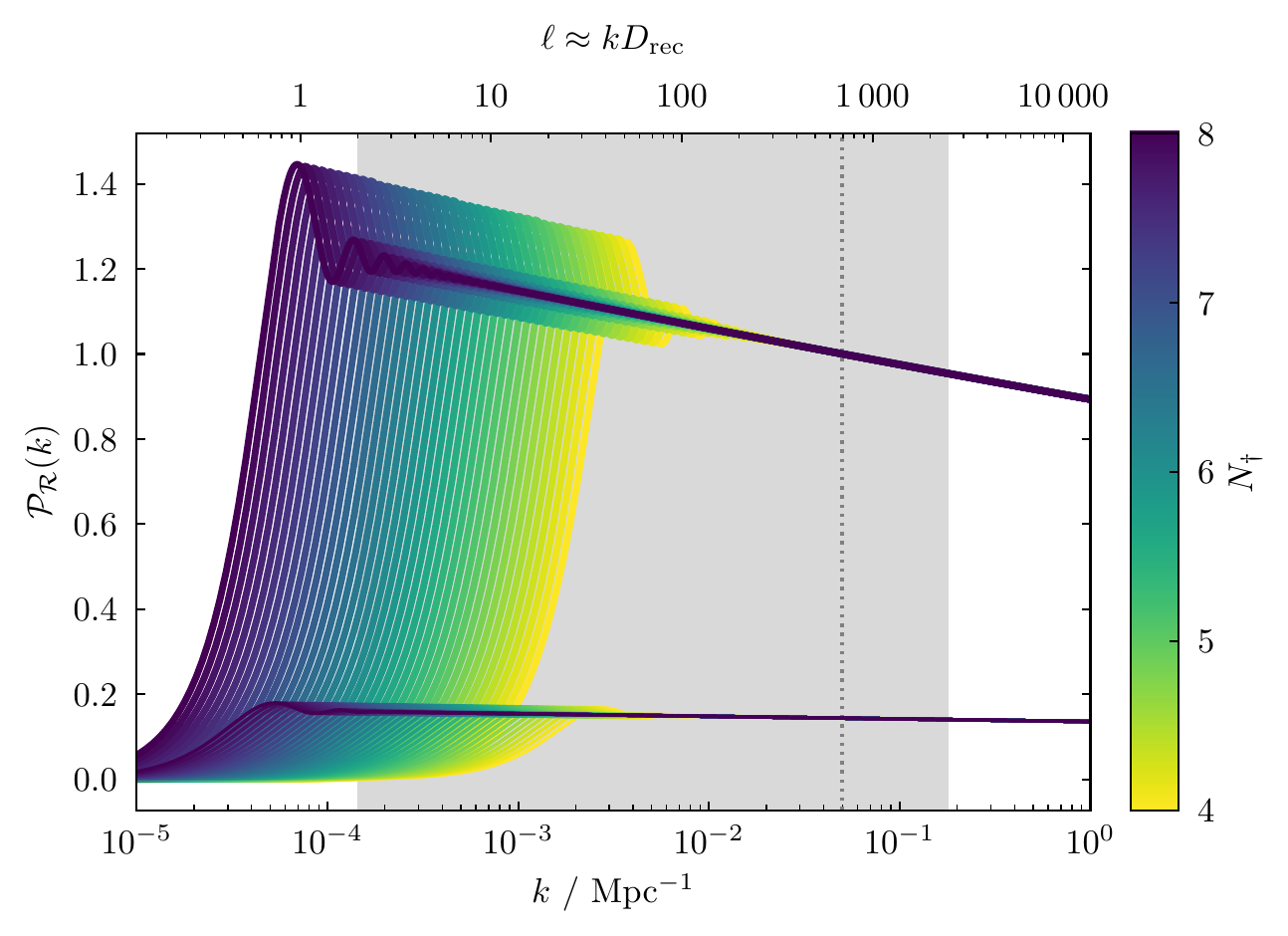}}    	
    \caption{PPS for scalar (upper) and tensor (lower) perturbations. 
    The left plot varies the number of observable \si{\efolds}~$N_\ast$ while keeping a fixed value of $N_\dagger=\SI{6}{\efolds}$. As denoted in \cref{eq:m2phi2_ns_r,eq:starobinsky_ns_r}, $N_\ast$ governs the spectral index~$n_\mathrm{s}$, i.e.\ the slope of the power spectrum, and the tensor-to-scalar ratio~$r$. The larger~$N_\ast$ the larger also~$n_\mathrm{s}$ and thus a smaller slope (closer to scale invariance). The tensor-to-scalar ratio on the other hand decreases with increasing~$N_\ast$. 
    The right plot varies the number of \si{\efolds}~$N_\dagger$ before horizon exit for a fixed value of $N_\ast=\SI{55}{\efolds}$. Both~$n_\mathrm{s}$ and~$r$ stay unaffected in this case. $N_\dagger$ instead governs the low-$k$ cutoff position of the PPS pushing it to ever smaller $k$-values as~$N_\dagger$ grows.}
    \label{fig:efolds}
\end{figure*}

For the evolution of the primordial perturbations we work directly with the primordial curvature perturbations~$\mathcal{R}$ and the tensor perturbations~$h$ as functions of cosmic time and for a given mode~$k$~\cite{Hobson2006,Adams2001}:
\begin{equation}
    \label{eq:scalarmodes}
	\ddot{\mathcal{R}}_k + \left( \frac{\dot\phi^2}{\si{\planckmass\squared}H} + \frac{2\ddot\phi}{\dot\phi} + 3H \right) \dot{\mathcal{R}}_k + \frac{k^2}{a^2} \mathcal{R}_k = 0 , 
\end{equation}
\begin{equation}
    \label{eq:tensormodes}
    \ddot{h}_k + 3 H \dot{h}_k + \frac{k^2}{a^2} h_k = 0 ,
\end{equation}
where the dot again refers to the derivative with respect to cosmic time.

For the numerical integration of the differential equations we loosely follow the scheme outlined in~\cite{Adams2001,Mortonson2009,Mortonson2010}. We reduce the differential equations into a first-order system and superimpose two orthogonal solutions. We start out by only evolving the background \cref{eq:background1,eq:background2,eq:eom}. At the start of inflation we start the integration of \cref{eq:scalarmodes,eq:tensormodes} for \emph{all} modes~$k$. Note that this is different from e.g.~\cite{Adams2001,Mortonson2009,Mortonson2010}. For kinetic dominance initial conditions our modes~$k$ do not necessarily lie well within the comoving Hubble horizon $k \ll (aH)^{-1}$ (cf.\ \cref{fig:hubble}) as during kinetic dominance the comoving Hubble horizon is still growing until it reaches its maximum at the onset of inflation. Thus, one cannot simply start the mode evolution when it is 100th the scale of the Hubble horizon as in~\cite{Mortonson2010}. For slow-roll~(SR) initial conditions this only affects the computation speed and is otherwise irrelevant, but for kinetic dominance~(KD) initial conditions this is important. So instead, we start the evolution for all modes at the onset of inflation.

The initial conditions for the mode equations (note, these are not the same as the initial conditions for the inflaton, i.e.\ not SR or KD initial conditions) are set through the definition of the quantum vacuum. For SR initial conditions for the inflaton field, typically, the Bunch-Davies vacuum is chosen, which defines the quantum vacuum via Hamiltonian diagonalization. For KD initial conditions, on the other hand, the vacuum choice becomes relevant, see e.g.~\cite{Contaldi2003,Armendariz-Picon2003,Handley2016}. In this paper we limit ourselves to the Bunch-Davies vacuum, leaving the exploration of alternative vacua to a later work.

We apply the Bunch-Davies vacuum on a linear combination of two orthogonal solutions.
The real and imaginary parts of the curvature perturbation~$\mathcal{R}_k(t)$ are plotted in \cref{fig:rk}, using SR and KD initial conditions for the inflaton respectively. For a good visualisation we use an inflaton mass of $m=\SI{5e-6}{\planckmass}$ for the quadratic potential and an amplitude of $\Lambda^2=\SI{e-5}{\planckmass}$ for the Starobinsky potential, and the mode $k=\SI{0.03}{\per\mega\parsec}$. Higher $k$-values would result in increasingly more oscillations.

We read off the frozen values of the primordial perturbations after horizon exit and obtain the scalar and tensor power spectra
\begin{align}
    \label{eq:scalarpower}
	\mathcal{P_R}(k) &= \frac{k^3}{2\pi^2} \left| \mathcal{R}_k \right|^2 , \\
    \label{eq:tensorpower}
	\mathcal{P}_\mathrm{t}(k) &= 2 \cdot \mathcal{P}_h(k)  = 2 \cdot \frac{k^3}{2\pi^2} \left| h_k \right|^2 , 
\end{align}
where the factor 2 in the tensor spectrum comes from the two possible polarization states of gravitational waves.

In order to compare our results to CMB data, we need to calibrate the perturbation scales. Calculations of the evolution of the universe from the end of inflation until today constrain the (observable) number of \si{\efolds} remaining during inflation after a given pivot scale~$k_\ast$ exited the Hubble horizon, to roughly within $50 \lesssim N_* \lesssim 60$~\cite{Liddle2003,Dodelson2003}. In accordance with Planck~\cite{Planck2015inflation} we choose $k_\ast=\SI{0.05}{\per\mega\parsec}$ for our pivot scale. We then calibrate our $k$-axis by determining the value~$a_\ast H_\ast$ (cf.\  \cref{fig:hubble}) for which~$N_\ast~\si{\efolds}$ of inflation remain after horizon exit
\begin{align}
	\label{eq:calibration}
	k \mapsto \frac{aH}{a_\ast H_\ast} k_\ast . 
\end{align}

In \cref{fig:PPSnumeric} we have plotted the numerical solutions of the PPS for quadratic and Starobinsky inflation, each with SR and KD initial conditions for the inflaton. In agreement with \cref{eq:m2phi2_ns_r,eq:starobinsky_ns_r} quadratic and Starobinsky inflation show a very similar spectral index~$n_\mathrm{s}$ and a tensor-to-scalar ratio~$r$ differing by about two orders of magnitude. 
As expected, the choice of SR or KD initial conditions does not affect small scales, since smaller scales freeze out later in the inflationary history when the slow-roll approximation is fully applicable for both cases. For larger scales we see oscillations and a cutoff towards small~$k$. 

The existence of the cutoff can be attributed to the preceding kinetically dominated phase and the brief period of fast-roll inflation~\cite{Boyanovsky2006,Boyanovsky2006a}. The larger modes spent less time within the horizon and the largest modes have actually \emph{never} been inside the horizon (scales greater than the maximum of the Hubble horizon in \cref{fig:hubble}).

The amplitude and frequency of the oscillations depend on the choice of the quantum vacuum, and consequently on the initial conditions for the curvature perturbations. Alternative choices for the quantum vacuum are proposed in~\cite{Armendariz-Picon2003,Handley2016}.

\subsection{\label{sec:efolds}Number of e-folds}

The exact position of the cutoff in the PPS for KD initial conditions depends on the initial value for~$\phi_\mathrm{p}$ in \cref{eq:phi0}. This is also related to the number of \si{\efolds} \emph{before} horizon crossing which we denote by~$N_\dagger$ as opposed to the \si{\efolds}~$N_\ast$ \emph{after} horizon crossing. Together they make up the total number of inflationary \si{\efolds}
\begin{align}
	N_\mathrm{tot} \equiv \ln\left(\frac{a_\mathrm{end}}{a_\mathrm{start}}\right) = N_\dagger + N_\ast . 
\end{align}
It is very hard to \emph{a-priori} constrain the total number of \si{\efolds}~$N_\mathrm{tot}$. Assuming inflation started after the Planck epoch, an upper bound on~$N_\mathrm{tot}$ can be set. For a quadratic potential with a roughly realistic inflaton mass of $m=\SI{5e-6}{\planckmass}$ such a bound is of an order of about $\max(N_\mathrm{tot}) \sim \SI{e10}{\efolds}$~\cite{Remmen2014}. 

Assuming the inflaton underwent a kinetically dominated phase before inflation, i.e.\ where $\dot\phi\gg V(\phi)$, we expect a significantly smaller number of \si{\efolds}, $N_\mathrm{tot}\ll\SI{e10}{\efolds}$. Stronger claims on the total amount of inflation have been made in the context of ``finite inflation''~\cite{Banks2003,Phillips2015} or ``just enough inflation''~\cite{Ramirez2012,Ramirez2012a,Cicoli2014}, where $N_\mathrm{tot}\gtrsim N_\ast$. Also, the expected amount of inflation can drop significantly depending on the choice of potential. While $\langle N_\mathrm{tot} \rangle \sim \num{e10}$ for the quadratic potential, it can turn out to be as low as $\langle N_\mathrm{tot} \rangle \sim \num{e1}\text{ or }\num{e2}$ for natural inflation depending on the symmetry breaking parameter~$f$ as shown in~\cite{Remmen2014}.

\Cref{fig:efolds} shows the effect of~$N_\dagger$ and~$N_\ast$ on the primordial power spectrum~$\mathcal{P}_\mathcal{R}(k)$ for the quadratic potential. The behaviour is very similar for the Starobinsky potential with the major difference being a significantly smaller tensor-to-scalar ratio for the Starobinsky model as can already be inferred from \cref{eq:m2phi2_ns_r,eq:starobinsky_ns_r}. As those equations suggest, we find that~$N_\ast$ governs both the spectral index~$n_\mathrm{s}$ and the tensor-to-scalar ratio~$r$. On the other hand, $N_\dagger$ leaves both these parameters invariant. Instead it shifts the cutoff position along the $k$-axis. More total \si{\efolds}~$N_\mathrm{tot}$, i.e.\ a longer period of inflation, and thus a larger~$N_\dagger$ pushes the cutoff to ever smaller $k$-values (larger scales). Thus, large scale CMB data will help us to constrain~$N_\dagger$ and~$N_\mathrm{tot}$.

Note that for SR initial conditions there is no clear start to inflation. One may therefore consider SR to correspond to the $N_\dagger, N_\mathrm{tot} \to \infty$ limit of KD.

\section{\label{sec:cmb}CMB power spectrum}

\begin{figure*}[tb]
	\centering
    \subfloat{\includegraphics[scale=0.9]{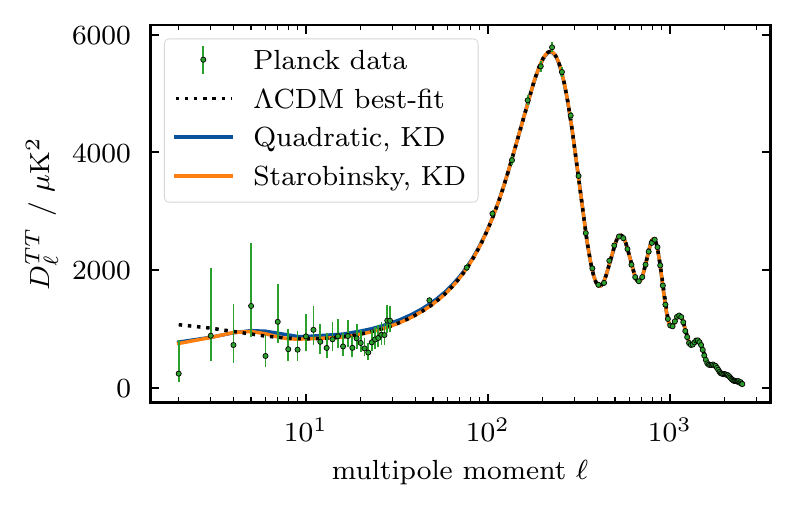}}
    \hfill
    \subfloat{\includegraphics[scale=0.9]{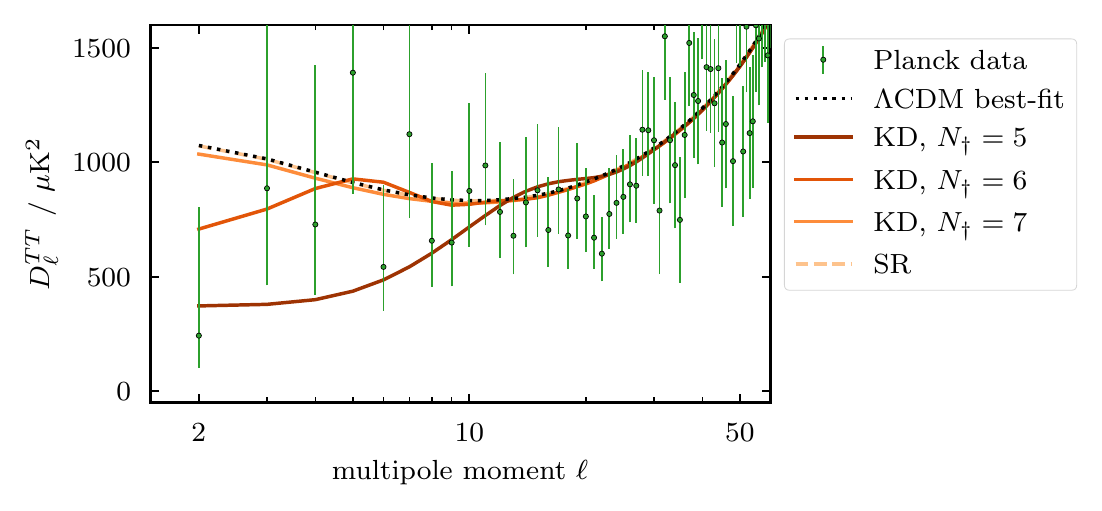}}    	
    \caption{CMB angular power spectrum $\mathcal{D}_\ell^{TT}\equiv\ell(\ell+1)\mathcal{C}_\ell^{TT}/(2\pi)$: The left hand side shows the best-fit lines (obtained from the MCMC analysis in \cref{sec:cmb}) for the $\Lambda$CDM model, and the quadratic and Starobinsky model with kinetic dominance~(KD) initial conditions respectively. The right hand side zooms in on the low-$\ell$ region and shows how the cutoff and oscillations from the PPS have translated through to the CMB power spectrum, where the cutoff position still depends on the value of~$N_\dagger$. Large values become indistinguishable to the slow-roll~(SR) case as the cutoff moves out of the observable region. 
}
    \label{fig:cmb}
\end{figure*}

To translate the primordial power spectra (PPS) from \cref{eq:scalarpower,eq:tensorpower} through to the angular power spectrum of the cosmic microwave background (CMB) we make use of the Boltzmann solver \texttt{CAMB}~\cite{Camb,Camb1,Camb2,Camb3}, which we modify such that it takes our PPS. To that end we first modify our input PPS such that they are normalised at the pivot scale~$k_\ast$ and the desired amplitude is then given by the \texttt{CAMB} parameter~$A_\mathrm{s}$
\begin{align}
	\mathcal{P_R}(k) \mapsto A_\mathrm{s} \cdot \frac{\mathcal{P_R}(k)}{\mathcal{P_R}(k_\ast)} . 
    \label{eq:ppsnormalization}
\end{align}
We can do this, because the background \cref{eq:background1,eq:background2,eq:eom} are invariant under a simultaneous rescaling of the time coordinate and the inflaton potential
\begin{equation} \label{eq:rescaling}
    t \mapsto {\sigma}^{-1}{t} , \:
    V(\phi) \mapsto \sigma^2 V(\phi)
    \Rightarrow\mathcal{P_R}(k) \mapsto \sigma^2 \mathcal{P_R}(k),
\end{equation}
effectively making a substitution of $\sigma^2 = 1/m^2$ or $\sigma^2=1/\Lambda^4$ to get a PPS~$\mathcal{P_R}(k,\sigma)$ independent of the potential amplitude. The PPS amplitude can then be linked to any desired mass~$m_0$ or amplitude~$\Lambda_0$ through $A_\mathrm{s}=m_0^2 \mathcal{P_R}(k,\sigma)$. 
The same results are obtained using the alternative Boltzmann solver \texttt{Class}~\cite{Class,Class1,Class2,Class3,Class4,Class5,Class6}.

\Cref{fig:cmb} shows the CMB angular temperature power spectrum for the Planck data~\cite{Planck}, for its $\Lambda$CDM best-fit model~\cite{Planck2015parameters}, and for the quadratic inflation model with kinetic dominance~(KD) initial conditions. The characteristic features of the KD initial conditions: low-$k$ cutoff and oscillation, are still apparent although diluted from convolution with the transfer functions. As for the PPS, the cutoff position depends on the number of \si{\efolds} before horizon exit~$N_\dagger$. For a sufficiently small value, the cutoff sinks into the low-$\ell$ lack of power found in the Planck data. The oscillations, however, are too heavily smoothed to follow the dip at multipoles $\ell$ at approximately 20--25. This is in line with the findings in~\cite{Scacco2015}.

\section{\label{sec:mcmc}MCMC analysis}



\begin{table*}[tbp]
\renewcommand{\arraystretch}{1.5}
\begin{ruledtabular}
\caption{\label{tab:params1} Marginalized parameter values at \SI{68}{\percent}~limits for different models and using CMB temperature data with low-$\ell$ polarization (TT+lowP). For the number of e-folds~$N_\dagger$ before and~$N_\ast$ after horizon exit we additionally provide the best-fit values as the data does not clearly delimit these parameters and they differ considerably from the \SI{68}{\percent}~limits. The posterior distribution for~$N_\dagger$ is essentially flat causing the mean to fall roughly in the middle of the defined prior range. However, there is a peak at small values causing the different best-fit value (cf.\ \cref{fig:bctHAnrNN}). $N_\ast$~is driven to high values for the quadratic model due to its correlation with the tensor-to-scalar ratio~$r$, and rather unconstrained for the Starobinsky model.}
\begin{tabular*}{\textwidth}{ l @{\hskip 0.6cm} cc @{\hskip 0.6cm} cc @{\hskip 0.6cm} c @{\hskip 0.6cm} c @{\hskip 0.6cm} c }
Parameters & \multicolumn{2}{c}{$N_\dagger$} & \multicolumn{2}{c}{$N_\ast$} & $n_\mathrm{s}$ & $r$ & $\ln(10^{10} A_\mathrm{s})$ \\
Prior ranges & \multicolumn{2}{c}{$[4, 15]$} & \multicolumn{2}{c}{$[50, 60]$} & $[0.885, 1.040]$ & $[0, 1]$ & $[2.5, 3.7]$ \\
TT+lowP & \SI{68}{\percent} limits & best-fit & \SI{68}{\percent} limits & best-fit & \SI{68}{\percent} limits & \SI{68}{\percent} limits & \SI{68}{\percent} limits \\
\hline
$\Lambda$CDM    &                     &             &          &          & $0.9655\pm 0.0063           $ & $                             $ & $3.089\pm 0.037$ \\
$r\Lambda$CDM   &                     &             &          &          & $0.9665\pm 0.0062           $ & $< 0.0504                     $ & $3.086\pm 0.036$ \\
Quadratic, SR &                     &             & $> 55.4$ & $60.00 $ & $0.9641^{+0.0022}_{-0.00069}$ & $0.1425^{+0.0027}_{-0.0086}   $ & $3.069\pm 0.032$ \\
Quadratic, KD & $9.8^{+3.4}_{-4.4}$ & $6.02     $ & $> 55.4$ & $57.48 $ & $0.9642^{+0.0022}_{-0.00067}$ & $0.1407^{+0.0026}_{-0.0085}   $ & $3.071\pm 0.032$ \\
Starobinsky, SR &                     &             &  ---     & $57.98 $ & $0.9644^{+0.0028}_{-0.0013} $ & $0.00365^{+0.00026}_{-0.00055}$ & $3.088\pm 0.033$ \\
Starobinsky, KD & $> 7.92           $ & $6.09     $ &  ---     & $57.07 $ & $0.9649^{+0.0027}_{-0.0011} $ & $0.00356^{+0.00021}_{-0.00053}$ & $3.089\pm 0.034$ \\
\end{tabular*}
\bigskip
\caption{\label{tab:params2}Marginalized parameter values at \SI{68}{\percent}~limits for different models and using temperature data with low-$\ell$ polarization (TT+lowP). Note, how the parameter values stay relatively similar across different models while the errors go down for models with explicit inflationary models (quadratic and Starobinsky) which can also be seen in the narrower contours in \cref{fig:bctHAnrNN}. This is attributed to the prior on~$N_\ast$ setting an effective, very narrow prior on the spectral index~$n_\mathrm{s}$ and the tensor-to-scalar ratio~$r$. Comparing slow-roll~(SR) and kinetic dominance~(KD) models, we find that these parameters do not distinguish between them at all.}
\begin{tabular*}{\textwidth}{ l c c c c c }
Parameters & $\Omega_b h^2$ & $\Omega_c h^2$ & $\tau$ & $H_0$ \\
Prior ranges & $[0.019, 0.025]$ & $[0.095, 0.145]$ & $[0.01, 0.40]$ & $[1.03, 1.05]$ on $100\theta_\mathrm{MC}$ \\
TT+lowP & \SI{68}{\percent} limits & \SI{68}{\percent} limits & \SI{68}{\percent} limits & \SI{68}{\percent} limits \\
\hline
$\Lambda$CDM    & $0.02223\pm 0.00023$ & $0.1197\pm 0.0022$ & $0.078\pm 0.019$ & $67.3\pm 1.0  $ \\
$r\Lambda$CDM   & $0.02224\pm 0.00023$ & $0.1195\pm 0.0022$ & $0.076\pm 0.019$ & $67.42\pm 0.98$ \\
Quadratic, SR & $0.02215\pm 0.00019$ & $0.1203\pm 0.0013$ & $0.067\pm 0.016$ & $67.02\pm 0.57$ \\
Quadratic, KD & $0.02216\pm 0.00019$ & $0.1202\pm 0.0013$ & $0.068\pm 0.016$ & $67.04\pm 0.57$ \\
Starobinsky, SR & $0.02222\pm 0.00019$ & $0.1201\pm 0.0013$ & $0.076\pm 0.016$ & $67.18\pm 0.56$ \\
Starobinsky, KD & $0.02222\pm 0.00019$ & $0.1199\pm 0.0013$ & $0.077\pm 0.017$ & $67.24\pm 0.57$ \\
\end{tabular*}
\end{ruledtabular}
\end{table*}

\begin{figure*}[p!]
	\centering
	\includegraphics[width=\textwidth]{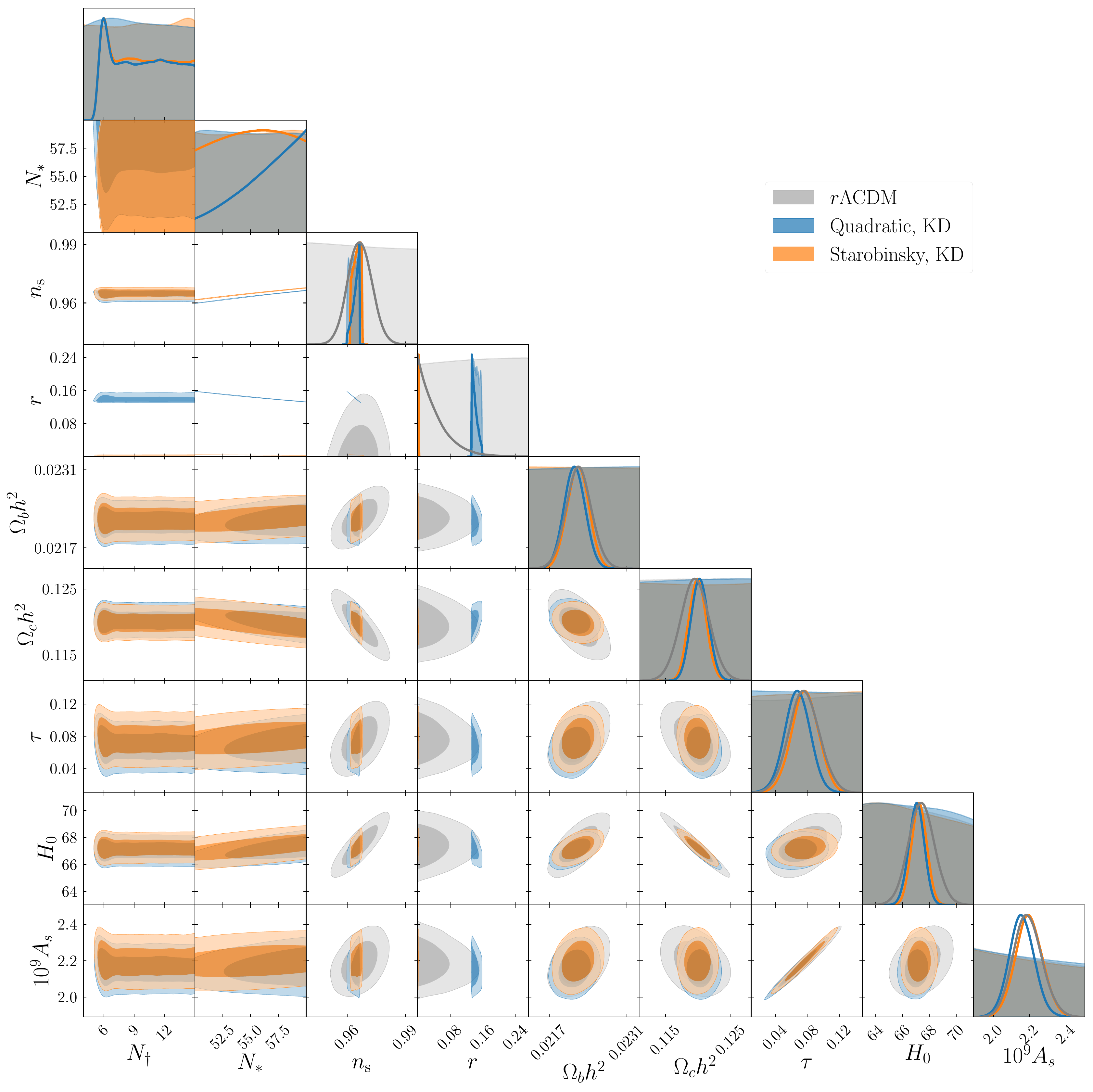}
    \caption{Triangle plot of the parameters: The number of \si{\efolds}~$N_\dagger$ before and~$N_\ast$ after horizon exit, the spectral index~$n_\mathrm{s}$, the tensor-to-scalar ratio~$r$, the baryon density parameter~$\Omega_\mathrm{b}$, the mass density parameter~$\Omega_\mathrm{c}$, the optical depth~$\tau$, the Hubble parameter~$H_0$, and the primordial amplitude of scalar perturbations~$A_\mathrm{s}$. The contours delimit the 0.68, and 0.95 levels. The shaded regions in the 1D plots on the diagonal correspond to the flat priors on the input parameters or the derived parameters in the case of~$n_\mathrm{s}$ and~$r$. The thicker solid lines in the 1D plots are the posterior distributions. \Cref{fig:nr} shows the $n_\mathrm{s}$-$r$-plane separately and enlarged. For the mean and standard deviation of marginalized parameters see \cref{tab:params1,tab:params2}.}
    \label{fig:bctHAnrNN}
\end{figure*}

We performed a Markov chain Monte Carlo (MCMC) analysis to extract the cosmological parameters of extended $\Lambda\mathrm{CDM}$ models alongside the kinetic dominance~(KD) initial conditions. To that end we used \texttt{CAMB}'s MCMC extension \texttt{CosmoMC}~\cite{CosmoMC,CosmoMC1,CosmoMC2} in conjunction with Planck's temperature and low-$\ell$ polarization data (TT+lowP) and corresponding likelihood code~\cite{Planck}. Additionally we perform a model comparison using \texttt{CosmoChord} which is a \texttt{PolyChord}~\cite{PolyChord,PolyChord1,PolyChord2} plug-in for  \texttt{CosmoMC}. \texttt{PolyChord} is a Bayesian inference tool for the simultaneous calculation of evidences and sampling of posterior distributions, and allows us to calculate the Bayes' factor of models. It performs well even on moderately high-dimensional posterior distributions, and can cope with arbitrary degeneracies and multi-modality. As such it is the successor to \texttt{MultiNest}~\cite{MultiNest,MultiNest1,MultiNest2,MultiNest3}, a variation of Nested Sampling~\cite{Skilling04}.

For our parameter estimation we added~$N_\dagger$ and~$N_\ast$ as new parameters in place of $n_s$. We put a flat prior within the range of $50<N_\ast<60$ in accordance with the expected number of observable \si{\efolds}~\cite{Dodelson2003,Liddle2003,Alabidi2006,Planck2015inflation}. For~$N_\dagger$ we chose a range from 4 to 15. We choose to cut values greater than $N_\dagger=\SI{15}{\efolds}$ as the PPS becomes observationally indistinguishable from the slow-roll~(SR) case. We retained the amplitude parameter~$A_\mathrm{s}$ to multiply our normalized PPS by, as already detailed in \cref{eq:ppsnormalization}. With these three parameters in place, the PPS is fully parametrised. Both the spectral index~$n_\mathrm{s}$ and the tensor-to-scalar ratio~$r$ turn into derived parameters inferred from the input PPS (cf.\ \cref{fig:efolds}). The remaining standard cosmological parameters were varied as for the $\Lambda$CDM case, namely the baryon density parameter~$\Omega_\mathrm{b}h^2$, the mass density parameter~$\Omega_\mathrm{c}h^2$, the optical depth~$\tau$, and the ratio of the sound horizon to the angular diameter distance~$\theta_\mathrm{MC}$. \Cref{fig:bctHAnrNN} shows a triangle plot (created using \texttt{GetDist}~\cite{GetDist}) of all these parameters and \cref{tab:params1,tab:params2} list the means of the marginalised parameters and their uncertainties.

The models considered are the standard $\Lambda$CDM model, $r\Lambda$CDM which is a one-parameter extension by the tensor-to-scalar ratio~$r$, and the quadratic and Starobinsky inflation models each with SR and KD initial conditions.

\subsection{Posteriors and priors on model parameters}

\begin{figure}[tb]
	\centering
    \includegraphics[width=\columnwidth]{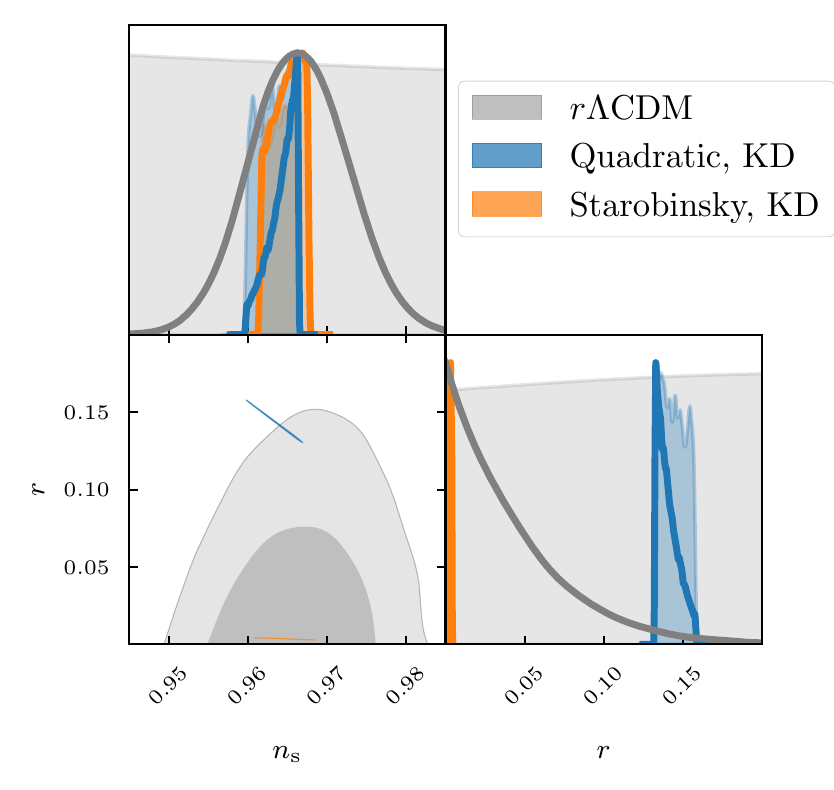}
    \caption{\label{fig:nr}Spectral index~$n_\mathrm{s}$ vs tensor-to-scalar ratio~$r$ triangle plot (zoomed in from \cref{fig:bctHAnrNN}). The shaded regions in the 1D plots denote the flat input prior for $r\Lambda$CDM and the derived priors for the inflationary models. The thick solid lines are the posteriors.}
\end{figure}

We begin by considering the constraints on the spectral index~$n_s$ and tensor-to-scalar ratio~$r$, detailed in the third and forth rows and columns of \cref{fig:bctHAnrNN}, and highlighted in \cref{fig:nr}. We plot only the $r\Lambda$CDM model and the Quadratic and Starobinsky model with KD initial conditions in \Cref{fig:bctHAnrNN,fig:nr} as the $\Lambda$CDM model and the SR inflation models are visually very similar to their counterparts for the shared parameters. The major difference lies in the additional parameters: the tensor-to-scalar ratio~$r$ for $r\Lambda$CDM and $N_\dagger$ for the KD inflation models.

As expected, we find significant differences for the amount of tensor modes, as~$r$ is significantly larger for the quadratic inflation model than for the Starobinsky model, and larger even than the \SI{68}{\percent} upper bound of the $r\Lambda$CDM model. 

Both inflationary models exhibit cut-off effects in their posterior contours. This is due to the relationship between~$N_\ast$, $n_\mathrm{s}$ and~$r$ from \cref{eq:m2phi2_ns_r,eq:starobinsky_ns_r}. The flat prior on~$N_\ast$ leads to an induced prior on~$n_\mathrm{s}$ and~$r$ that is much narrower than the traditional $\Lambda$CDM or $r\Lambda$CDM priors. This constraint is then projected onto the other parameters.  

Given this \emph{a-priori} predictivity in~$r$ and~$n_s$, one might object at this point that the prior range chosen for~$N_\ast$ is too narrow. However, the broad prior ranges for~$n_s$ and~$r$  in the $r\Lambda$CDM model may be viewed as a phenomenological model-averaging over a wide class of inflationary models. It allows $r\Lambda$CDM to represent and compare many inflation models in an $n_\mathrm{s}$-$r$-plot (\cref{fig:nr}). Thus, it is only natural that specific models give narrower priors on parameters such as the spectral index or the tensor-to-scalar ratio, and it is this which eventually allows the falsification of different inflationary models.

Consider now the marginalised posteriors involving the number of \si{\efolds} before and after horizon exit ($N_\dagger$ and $N_\ast$), detailed in the first and second rows and columns of \cref{fig:bctHAnrNN}, and best-fit values in \cref{tab:params1,tab:params2}.

Neither $N_\ast$ nor $N_\dagger$ are clearly constrained for either model. For quadratic inflation, $N_\ast$ is driven to high values in order to decrease the tensor-to-scalar ratio~$r$ and thus we get a lower bound for the \SI{68}{\percent}~limits. For the Starobinsky model, $N_\ast$ is essentially only constrained through the prior choice which was here taken to be $50<N_\ast<60$. A small amount of constraining power comes from the correlation with~$n_\mathrm{s}$. 
$N_\dagger$ on the other hand behaves very similarly for both inflation models. While very low values are clearly ruled out by the data,  the posterior plateaus for larger values, the exception being a single peak at about $N_\dagger=6$ roughly a factor 2 above the plateau. Low values will push the power spectrum cutoff unfavourably far into the data. The best-fit value manages to position the cutoff such that it aligns with the low-$\ell$ lack of power. Once the cutoff is pushed out of the observable region, KD is equivalent to SR, there is no change to the CMB power spectrum, and all large values of~$N_\dagger$ become equally likely.

Finally, from the remaining rows and columns of \cref{fig:bctHAnrNN}, and \cref{tab:params1,tab:params2} one can see that all of the standard cosmological parameters ($\Omega_\mathrm{b}h^2$, $\Omega_\mathrm{c}h^2$, $\tau$, $H_0$, $A_\mathrm{s}$, $n_s$) for all additional models are consistent with the values obtained for the standard $\Lambda$CDM model.

\subsection{Evidences}

\begin{figure}[tb]
	\includegraphics[width=\columnwidth]{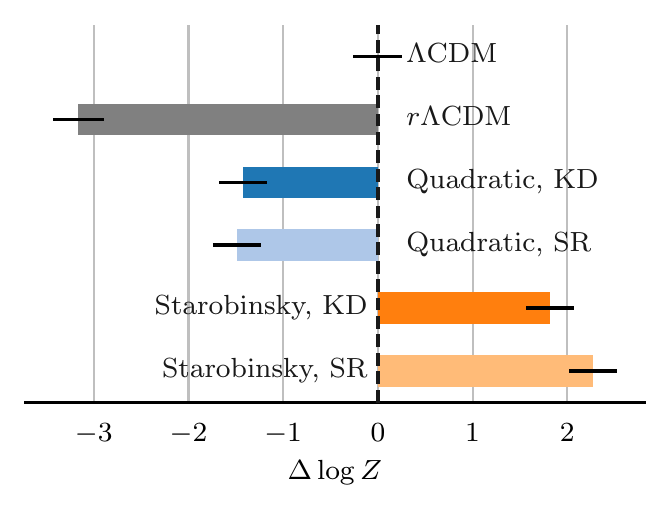}
    \caption{\label{fig:evidences}Difference of log evidences~$\Delta\log Z$ with respect to the $\Lambda$CDM model as reference. The errors are roughly a quarter log unit throughout.}
\end{figure}

From \cref{fig:nr} we already judged the Starobinsky model to perform better than the quadratic model, since the line for the Starobinsky model sits in the middle of the $r\Lambda$CDM contour, whereas the line for the quadratic model lies on the outer edge of the \SI{95}{\percent} contour. For a proper model comparison we calculate and compare their respective Bayesian evidences. Using \texttt{CosmoChord} we calculated the evidences $Z\equiv P(\mathcal{D}|\mathcal{M})$ for a given model~$\mathcal{M}$ using the Planck data~$\mathcal{D}$. \Cref{fig:evidences} visualizes the Bayes' factors, i.e.\ the difference of log evidences~$\Delta\log Z$ where we use the $\Lambda$CDM model as a reference model. The prior ranges used in the model comparison are listed in \cref{tab:params1,tab:params2}.

Comparing $\Lambda$CDM and $r\Lambda$CDM shows the effect a single additional parameter can have. Though $r\Lambda$CDM has an additional parameter and thus can make a greater variety of predictions, it also has to spread its predictive probability over a greater volume of parameter space and thus more thinly. This penalizes $r\Lambda$CDM considerably. For the comparison here we have chosen a prior range of $r\in [0, 1]$, which reads as an assumption that the tensor modes are smaller than the scalar modes.

As in the standard $\Lambda$CDM model, the SR models vary a total of six parameters. One of those parameters~$N_\ast$ replaces the spectral index~$n_\mathrm{s}$, which becomes a derived parameter (as does the tensor-to-scalar ratio~$r$). The inflation models with KD initial conditions introduce one additional input parameter~$N_\dagger$, resulting in a total of seven parameters.

As expected the quadratic model is disfavoured compared to the Starobinsky model with a difference of about 3 log units, mainly driven by the high tensor-to-scalar ratio~$r$ in the quadratic model. Due to their reduced parameter space, or equivalently their increased predictivity, they both outperform the very general $r\Lambda$CDM model. Only the Starobinsky model with its very low tensor modes manages to do better than the standard $\Lambda$CDM model, which effectively conditions $r$ to be zero. When comparing SR initial conditions to KD initial conditions the data do not show a clear preference towards one model or the other. Considering that an additional parameter is used for the KD case, the model manages to make up for the associated Occam penalty factor with a slightly better fit to the data.

\subsection{Power spectrum predictive posteriors and Kullback-Leibler divergences}

\begin{figure*}[tbp]
	\centering
    \null\hfill
    \subfloat[PPS of scalar~$\mathcal{P_R}(k)$ and tensor~$\mathcal{P}_\mathrm{t}(k)$ perturbations.]{\includegraphics[]{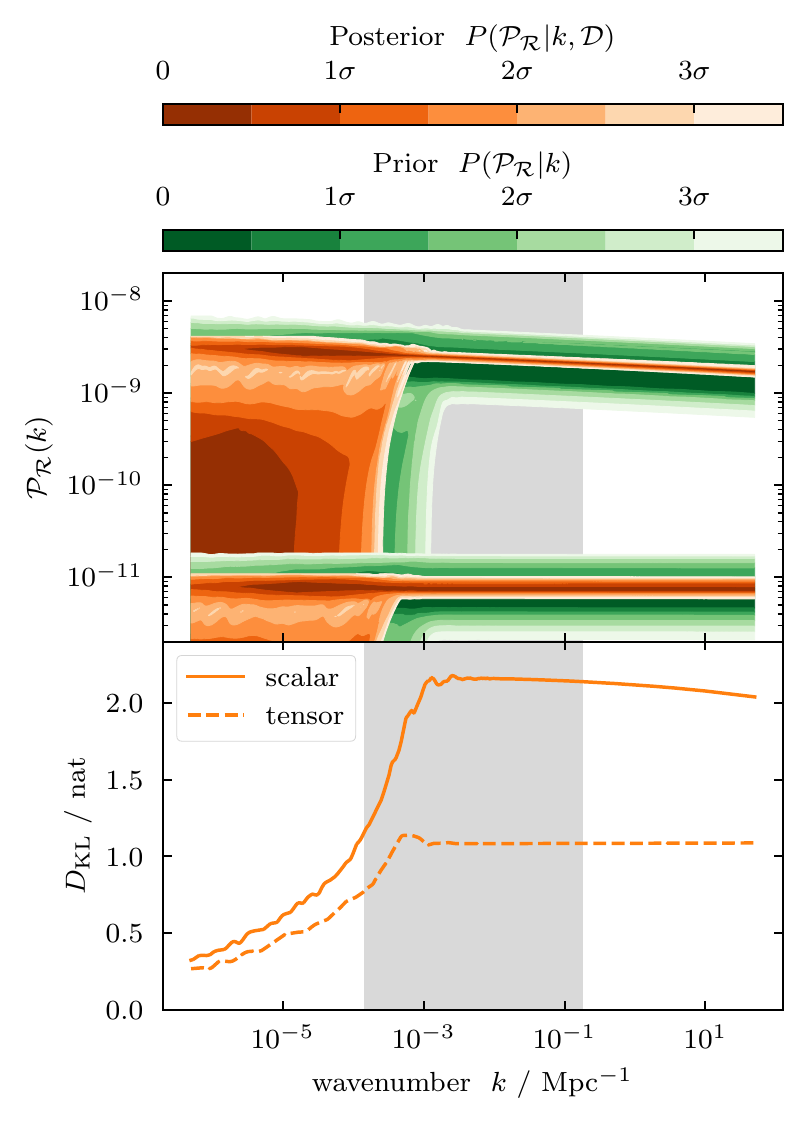}}
    \hfill
	\subfloat[CMB angular temperature power spectrum $\mathcal{D}_\ell^{TT}\equiv\ell(\ell+1)\mathcal{C}_\ell^{TT}/(2\pi)$.]{\includegraphics[]{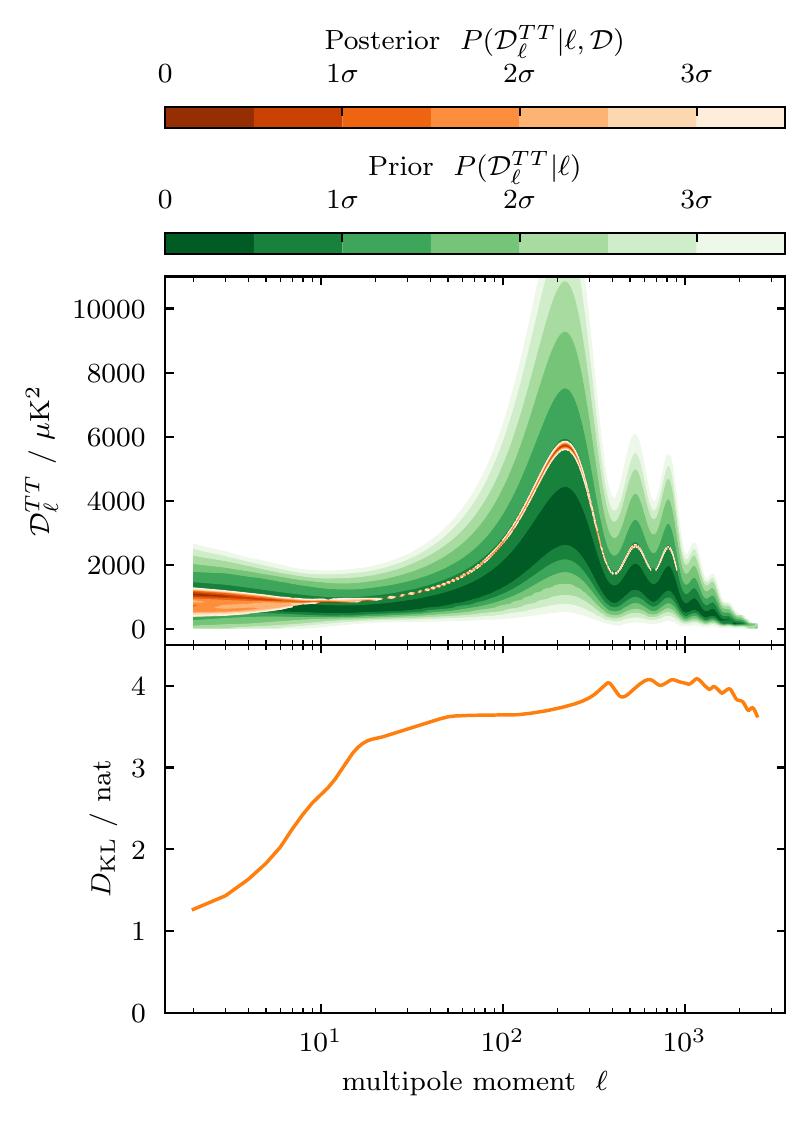}}
    \hfill\null
    \caption{\label{fig:density}The upper panels show density plots of the primordial power spectra~$\mathcal{P_R}$ and~$\mathcal{P}_\mathrm{t}$ (left) and the CMB temperature spectrum~$\mathcal{C}_\ell^{TT}$ (right) for parameter samples from the prior (green) and posterior (orange) distributions of the MCMC runs for the Starobinsky model. The lower panels show the corresponding plots for the relative entropy or Kullback-Leibler divergence~$D_\mathrm{KL}$ when going from the prior to the posterior distribution. The data~$\mathcal{D}$ is very constraining for large wavenumbers $k\gtrsim\SI{e-3}{\per\mega\parsec}$ and multipoles $\ell\gtrsim 10$, and drives up the information gain accordingly in those domains. From there the relative entropy plummets to roughly a fourth towards larger scales reflecting the lack of constraining power of the data. This is where the power spectrum cutoff can sink in. (Figure created using \texttt{fgivenx}~\cite{fgivenx}.)}
\end{figure*}

\begin{figure*}[tbp]
	\centering
    \null\hfill
    \subfloat[PPS of scalar~$\mathcal{P_R}(k)$ and tensor~$\mathcal{P}_\mathrm{t}(k)$ perturbations.]{\includegraphics[]{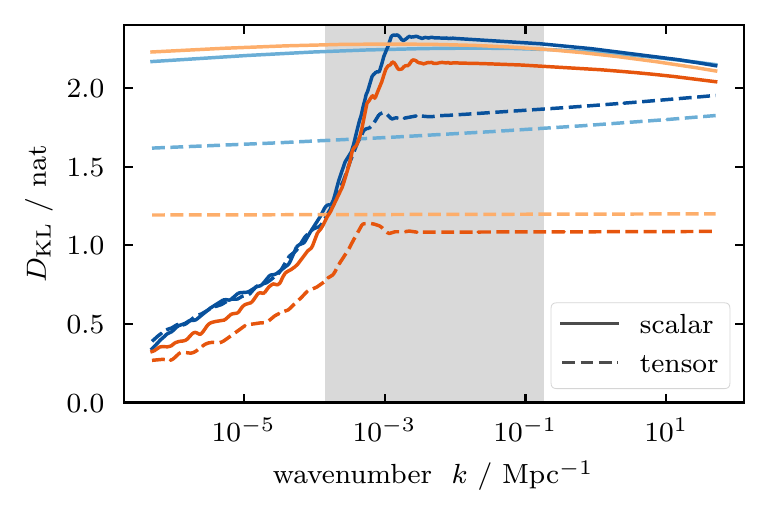}}
    \hfill
	\subfloat[CMB angular temperature power spectrum $\mathcal{D}_\ell^{TT}\equiv\ell(\ell+1)\mathcal{C}_\ell^{TT}/(2\pi)$.]{\includegraphics[]{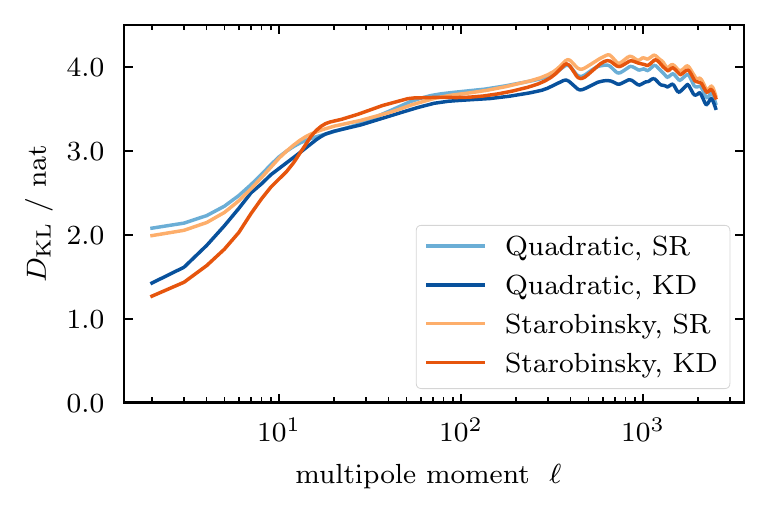}}
    \hfill\null
	\caption{\label{fig:dkl}Kullback-Leibler divergence comparing Quadratic vs Starobinsky inflation, and KD vs SR initial conditions.}
\end{figure*}

The major observable differences between the SR and the KD cases are the low-$\ell$ cutoff and oscillations in the power spectrum. In the upper half of \cref{fig:density} we show the prior and posterior densities of MCMC samples for both PPS and CMB power spectra for the Starobinsky model with KD initial conditions. The low-$k$ and low-$\ell$ cutoff from KD is not pushed out by the data but stays at the lower end of the observable region. We calculated the relative entropy or Kullback-Leibler divergence~$D_\mathrm{KL}$ going from the prior distribution to the posterior distribution (bottom plots in \cref{fig:density}). While the information gain throughout most of the spectrum is rather high and roughly constant, it drops off to roughly a fourth of its value towards the largest observable scales due to cosmic variance. 

\Cref{fig:dkl} additionally includes the divergence for the quadratic model and for SR initial conditions. The quadratic model shows a higher information gain than the Starobinsky model, which is most prominent for the tensor modes of the PPS. This is related to the tensor-to-scalar ratio being driven to small values. Assuming the quadratic model was the correct model, one knows that~$N_\ast$ would need to be high in order to get a sufficiently low tensor-to-scalar ratio. The higher information gain at large scales in case of SR initial conditions is attributed to the rigidity of the model. Assuming this time that SR initial conditions are correct, the data constrain the amplitude at small scales and the SR model then tells us that there must be a similar amplitude at large scales.

\section{\label{sec:conclusions}Conclusions}

We have shown that using kinetically dominated~(KD) initial conditions instead of slow-roll~(SR) initial conditions for homogeneous and isotropic single-field inflation causes oscillations and a cutoff towards large scales in the primordial power spectrum (PPS). The position of oscillations and cutoff is governed by the amount of inflation~$N_\dagger$ preceding horizon exit for any given pivot mode. The amount of inflation~$N_\ast$ after horizon exit determines the scalar spectral index~$n_\mathrm{s}$ and the tensor-to-scalar ratio~$r$.

We illustrate how these features carry through to the CMB power spectrum, where the cutoff in the PPS can sink into the low-$\ell$ lack of power in the CMB. The oscillations get washed out going from the PPS to the CMB such that they are not strong enough to model the dip in CMB power at multipoles~$\ell$ of approximately 20--25.

We perform an MCMC analysis and find that all standard cosmological parameters ($\Omega_\mathrm{b}$, $\Omega_\mathrm{c}$, $\tau$, $H_0$, $A_\mathrm{s}$, $n_\mathrm{s}$) for all the models taken into consideration ($r\Lambda$CDM, quadratic inflation with SR and KD initial conditions, Starobinsky inflation with SR and KD initial conditions) are consistent with the standard $\Lambda$CDM model. As expected, we find significant differences for the amount of tensor modes, favouring Starobinsky over quadratic inflation. Both the \si{\efolds}~$N_\dagger$ and~$N_\ast$ cannot be clearly estimated. The amount of inflation before horizon exit can be constrained from below and shows a peak at about $N_\dagger=\SI{6}{\efolds}$. From there it rapidly drops off to about half the peak amplitude and plateaus. This reflects that KD initial conditions are indistinguishable from SR initial conditions for large values of~$N_\dagger$. The amount of observable inflation~$N_\ast$ is essentially unconstrained, hence any constraints are mostly driven by the choice of prior. 

In a model comparison the Starobinsky model performs better and the quadratic model worse than the standard $\Lambda$CDM model. They both perform significantly better than the $r\Lambda$CDM model. Although we do not find a significant difference between the use of SR or KD initial conditions in terms of evidence, it is intriguing that the KD model manages to balance the penalty for an additional parameter with a slightly improved fit for small $N_\dagger$ at low multipoles, due to its effect on the overall power level in this region.

Finally, in an analysis of the posterior density and the Kullback-Leibler divergence, we confirm that most of the information gain from the data happens on small scales, i.e.\ for large multipoles. It will be interesting to consider in future work whether including large scale polarization data from a future cosmic variance limited CMB experiment can help to discriminate more definitively between SR and KR conditions in terms of their effects on low-$\ell$ CMB power.

\begin{acknowledgments}
This work was performed using the Darwin Supercomputer of the \href{http://www.hpc.cam.ac.uk/}{University of Cambridge High Performance Computing Service}, provided by Dell Inc.\ using Strategic Research Infrastructure Funding from the Higher Education Funding Council for England and funding from the Science and Technology Facilities Council, as well as resources provided by the \href{http://www.csd3.cam.ac.uk/}{Cambridge Service for Data Driven Discovery (CSD3)} operated by the University of Cambridge Research Computing Service, provided by Dell EMC and Intel using Tier-2 funding from the Engineering and Physical Sciences Research Council (capital grant EP/P020259/1), and \href{www.dirac.ac.uk}{DiRAC funding from the Science and Technology Facilities Council}.

LTH would like to thank the Isaac Newton Trust and the STFC for their support. WJH was supported by a Gonville \& Caius Research Fellowship.
\end{acknowledgments}

\bibliography{KineticDominanceBib,CollaborationBib,urlBib,footnote}

\end{document}